\documentclass[conference]{IEEEtran}
\IEEEoverridecommandlockouts
\usepackage{cite}
\usepackage{amsmath,amssymb,amsfonts}
\usepackage{graphicx}
\usepackage{textcomp}
\usepackage{xcolor}
\usepackage{comment}
\usepackage{url}
\usepackage{algorithm}
\usepackage{algpseudocode}

\def\BibTeX{{\rm B\kern-.05em{\sc i\kern-.025em b}\kern-.08em
    T\kern-.1667em\lower.7ex\hbox{E}\kern-.125emX}}
    
\newcommand{\feedback}[1]{\textbf{#1}}
\newcommand{\ncn}[1]{\feedback{NCN: #1}}

\newcommand{\ronak}[1]{\feedback{Ronak: #1}}

\begin{document}

\title{Intent-based Meta-Scheduling in Programmable Networks: A Research Agenda\thanks{Financial support for this work from a research grant from the Ministry of Electronics and Information Technology, Govt. of India, is gratefully acknowledged. The authors also wish to thank Anurag Kumar, Chandra R. Murthy, and Bharat Dwivedi for their comments.}}


\author{\IEEEauthorblockN{Nanjangud C.~Narendra, Ronak Kanthaliya, Venkatareddy Akumalla}
\IEEEauthorblockA{Dept. of Electrical Communication Engineering and FSID, Indian Institute of Science\\
Bangalore, India\\
ncnaren@gmail.com; ronak@fsid-iisc.in; venkatareddy@fsid-iisc.in}
}

\maketitle

\begin{abstract}

The emergence and growth of 5G and beyond 5G (B5G) networks has brought about the rise of so-called ``programmable'' networks, i.e., networks whose operational requirements are so stringent that they can only be met in an automated manner, with minimal/no human involvement. Any requirements on such a network would need to be formally specified via \emph{intents}, which can represent user requirements in a formal yet understandable manner. Meeting the user requirements via intents would necessitate the rapid implementation of resource allocation and scheduling in the network. Also, given the expected size and geographical distribution of programmable networks, multiple resource scheduling implementations would need to be implemented at the same time. This would necessitate the use of a \emph{meta-scheduler} that can coordinate the various schedulers and dynamically ensure optimal resource scheduling across the network. 

To that end, in this position paper, we propose a research agenda for modeling, implementation, and inclusion of intent-based dynamic meta-scheduling in programmable networks. Our research agenda will be built on \emph{active inference}, a type of causal inference. Active inference provides some level of autonomy to each scheduler while the meta-scheduler takes care of overall intent fulfillment. Our research agenda will comprise a strawman architecture for meta-scheduling and a set of research questions that need to be addressed to make intent-based dynamic meta-scheduling a reality.

\end{abstract}


\section{Introduction}\label{sec:intro}

The growth of programmable networks, driven by advances in 5G/6G technologies~\cite{haque2023survey}, has raised the need for rapid automated resource scheduling approaches. In particular, for 6G networks, resource scheduling is expected to be implemented within the sub-millisecond timeframe to meet 6G's stringent latency requirements. \emph{Intents\cite{niemoller2023autonomous,Peter,banerjee2021intent,mehmood2023intent}} are being seen as an effective mechanism for such rapid resource scheduling. Intents are at the same time human-understandable and machine-readable and are emerging as the standard approach for requirements specification and tracking in most telecommunication standards bodies such as 3GPP\footnote{\url{https://www.3gpp.org/}}, Telemanagement Forum\footnote{\url{https://www.tmforum.org/}}, and O-RAN\footnote{\url{https://www.o-ran.org/}}.

Within the programmable networks area, the trend is towards disaggregation of the network via architectures such as O-RAN~\cite{polese2022understanding}. One key feature of O-RAN relevant for us, is that it emphasizes separation of control and user planes in wireless networks. This separation enables the decomposition of intents from the user level down to the Radio Unit (RU) level, to facilitate optimal resource scheduling.

However, another key issue with programmable networks is their size and scale, which is expected to be much larger than the networks of today. Such networks are expected to be dense~\cite{23.501,38.300,38.413}, requiring special scheduling approaches tailored to dense networks~\cite{fulber2024genetic}. Furthermore, such large-scale networks could also be subdivided into administrative domains~\cite{christou} and hence, would need to be managed in a distributed manner. 

Hence the combination of ultra-low latency and large size and scale would make the job of resource scheduling extremely complex. The particular concern here would be the large number of scheduling algorithms that would need to be simultaneously implemented to cater to multiple user requests. This would increase the possibility of conflicts, necessitating the establishment of a \emph{meta-scheduling} approach~\cite{min2023meta} to coordinate among the schedulers.

In this position paper, we investigate this crucial research issue of meta-scheduling. We propose the use of intent-based management to model and implement meta-scheduling to coordinate and control the numerous schedulers that would be running in a programmable network. Capitalizing on the disaggregated nature of O-RAN, we show how intents can be decomposed from the user level, all the way down to the RU level to enable resource scheduling, and how this intent management hierarchy can be managed via meta-scheduling approaches. In particular, we show how the newly emerging technique of \emph{active inference}~\cite{sedlak2024active,casamayor2024deepslos}, derived from the well-established idea of \emph{causal inference}~\cite{pearl2016causal}, can help design and implement optimal meta-schedulers that can also facilitate hierarchical and federated learning approaches for scheduling~\cite{habib2023intent,erdol2022federatedmetalearningtrafficsteering}. In addition, we will present our research agenda in this space, which will comprise the key research questions to be addressed to make intent-based meta-scheduling a reality.

\section{Intent-driven Programmable Networks}\label{sec:disagg}

The key aspect of a programmable network is that control of the network is separated from operation. This lends itself to making the network programmable via \emph{intents}. An intent, as defined by the TeleManagement Forum (TMForum)~\cite{intent}, ``is the formal specification of all expectations including requirements, goals, and constraints given to a technical system''. From a user’s perspective, it expresses what a system is expected to achieve. It includes all the system needs to know, i.e., goals, requirements, constraints, etc. It needs formal modeling and common semantics to be understandable to the system; however, it is also intuitively understandable by humans. It is not only used on the human-machine interface, but in internal goal-setting between sub-systems, and this aspect makes intents valuable for our purposes, as will be seen later in this paper. Natural language and other domain-specific languages can be used which requires local interpretation and translation into the common intent model. It also has its life cycle actively managed by the intent creator through the intent API~\cite{intent,10255479}. Intent-based management has started to be used in several trial customer deployments, e.g.,~\cite{intent-operation}.

\begin{figure*}
        \begin{center}
                \includegraphics[scale=0.40]{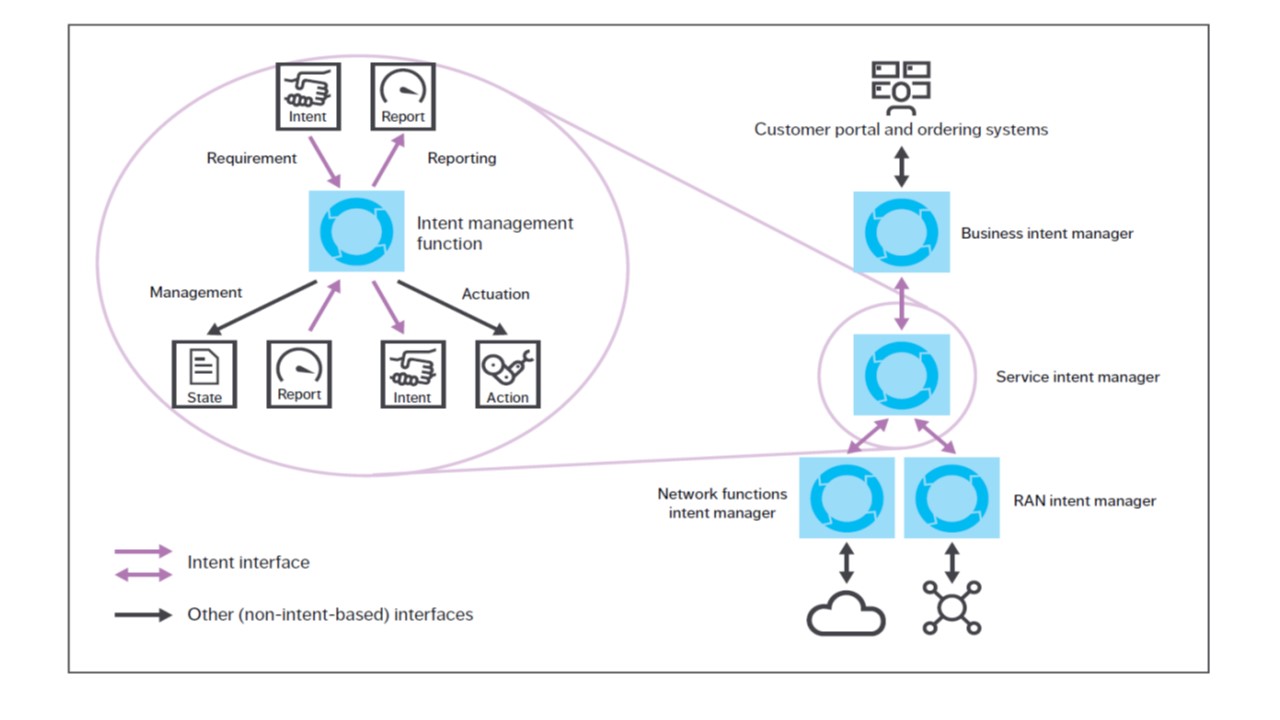}
        \end{center}
        \caption{Intent Decomposition - from~\cite{niemoller2023autonomous}}
        \label{fig:jorg}
\end{figure*}

The crucial aspect of intents is that they are decomposable. That is, as shown in Fig.~\ref{fig:jorg}, an intent can be specified at the Business Support System (BSS) layer, and decomposed further down to the Operations Support System (OSS), RAN, Transport, and Core layers of the network stack. This lends itself to the step-wise decomposition of user requirements at the BSS layer down to the network and 5G core layers. This property of intents makes them suitable for adaptable and flexible decision-making in programmable networks.

At each level in the network stack, intent management functions (IMF) can be defined, whose task is to translate the intents defined by IMFs in the upper layer into intents that IMFs can implement at the lower layer. An IMF would therefore be performing the role of an \emph{intent owner} or \emph{intent handler}, depending on when it is (respectively) assigning or decomposing intents. When the intent can no longer be decomposed at the lowest layer, the intent handler would have to implement the intent.

\begin{figure}
        \begin{center}
                \includegraphics[scale=0.28]{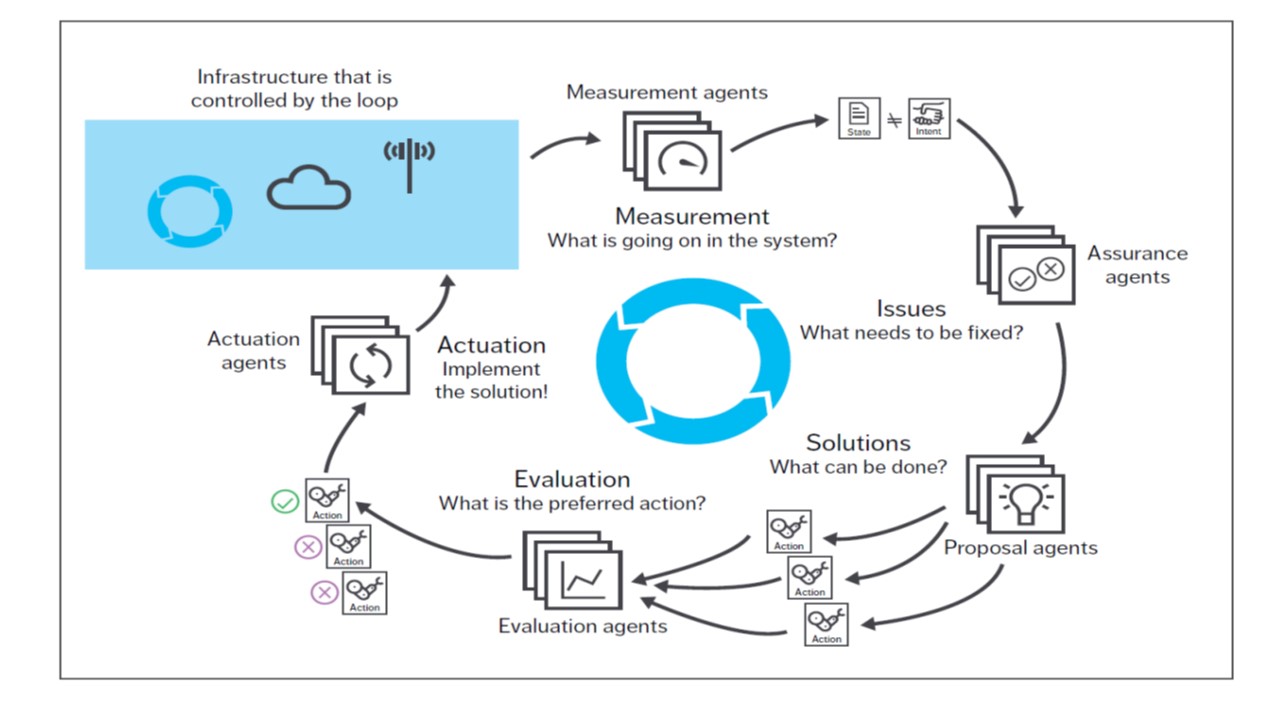}
        \end{center}
        \caption{Intent Management Loop - from~\cite{niemoller2023autonomous}}
        \label{fig:lifecycle}
\end{figure}

Each IMF would therefore go through a closed loop, driven by machine reasoning, as shown in Fig.~\ref{fig:lifecycle}. This comprises the following: (a) measurement agents that report the state of the network at the level below that of the IMF and report to it; (b) assurance agents, that determine what needs to be implemented based on the arriving intent; (c) proposal agent that proposes one or more intent decompositions, comprising a combination of decomposition and actuation as needed; (d) evaluation agent, that evaluates the proposals and selects the best one; and (e) actuation agent that implements the actual decomposition and actuation. Of course, all this is underpinned by a cognitive framework~\cite{DBLP:conf/aiml2/KattepurConrad} to operate the agents and help them perform machine reasoning tasks to achieve their respective objectives.

Automated intent decomposition, although still at a nascent stage, is emerging as an active research area. One example of intent decomposition, which also incorporates distribution of intents across administrative domains, is presented in~\cite{christou}. An energy-aware intent decomposition algorithm is presented in~\cite{wang2024networkintentdecompositionoptimization}. Intent decomposition using event calculus to represent the intents, and logical reasoning to model the intent decomposition process, is presented in~\cite{zhang2023intent}. Intent decomposition and propagation of the decomposed intents for network slice design is presented in~\cite{gritli}. 

By way of exposition, for readers unfamiliar with intent-based management, we have summarized the intent decomposition approach from~\cite{christou} in Appendix~\ref{sec:app-a}. We have chosen~\cite{christou} since it provides an overall perspective of intent decomposition, while also illustrating the highly distributed and multi-domain nature of programmable networks.

\section{Intent-based Resource Allocation and Scheduling in Programmable Networks}\label{sec:resalloc}

\begin{figure}
        \begin{center}
                \includegraphics[scale=0.28]{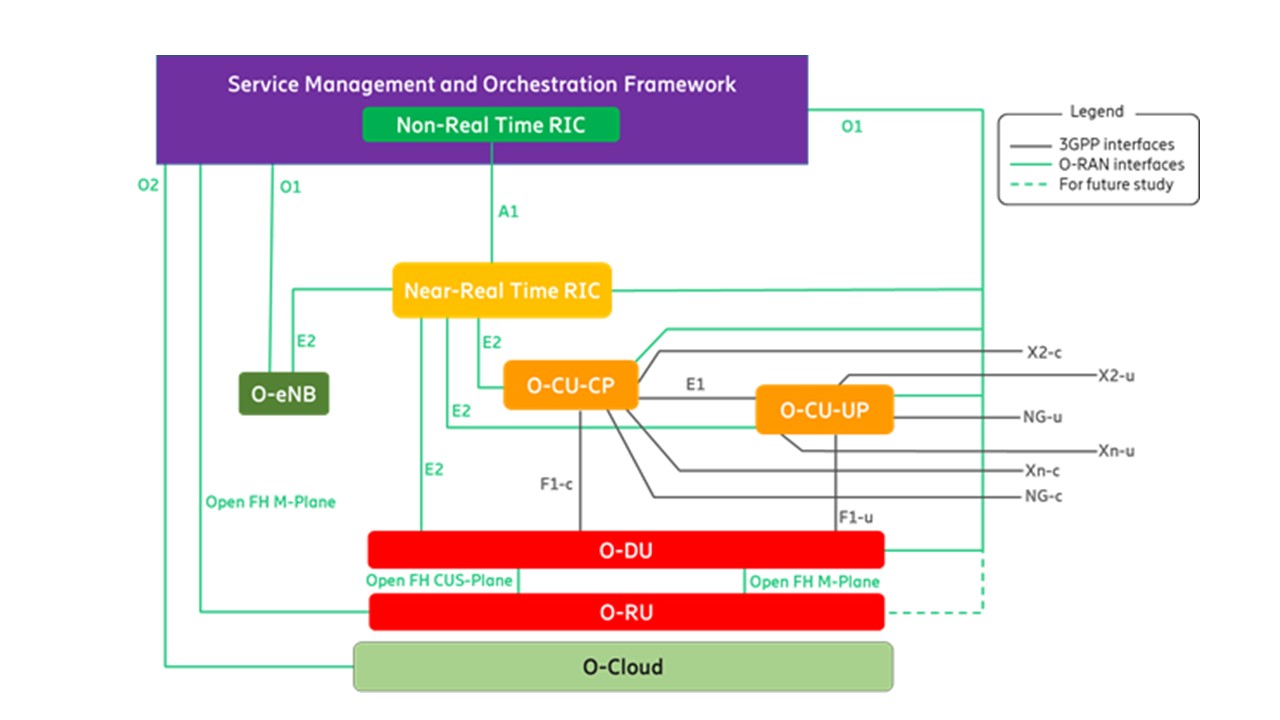}
        \end{center}
        \caption{O-RAN Logical Architecture}
        \label{fig:figure-02}
\end{figure}

We position our intent-based meta-scheduling approach within O-RAN~\cite{polese2022understanding}, since it is the latest version of a programmable network. Moreover, the disaggregated nature of O-RAN makes it suitable to incorporate and extend the intent decomposition approach depicted in Fig.~\ref{fig:jorg}. The O-RAN logical architecture is as depicted in Fig.~\ref{fig:figure-02}.

\begin{figure}
        \begin{center}
                \includegraphics[scale=0.42]{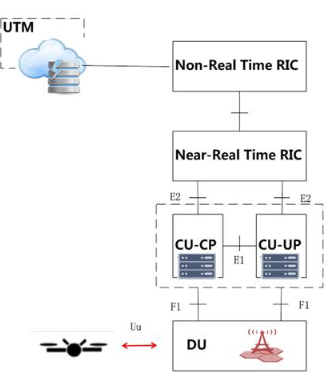}
        \end{center}
        \caption{Flight Path-based Dynamic UAV Radio Resource Allocation}
        \label{fig:figure-07}
\end{figure}

Consider, for example, the drone use case as shown in Fig.~\ref{fig:figure-07} (this has been reproduced from the O-RAN Use Case Analysis Report~\cite{o-ran-standards}). The drone has a network-layer connection to its nearest 5G cell, which is part of the zone of an edge site. The drone's connection to the 5G cells could be changed via network-layer handover, based on the radio resource management (RRM) algorithm employed at the Control Unit (CU-CP) of the O-RAN system that runs the network.


Context-based dynamic handover management for this use case will allow operators to adjust radio resource allocation policies through the O-RAN architecture, reducing latency and improving radio resource utilization. This would be done via the UTM (UAV Traffic Management) module as depicted in Fig.~\ref{fig:figure-07}.

\begin{figure}
        \begin{center}
                \includegraphics[scale=0.40]{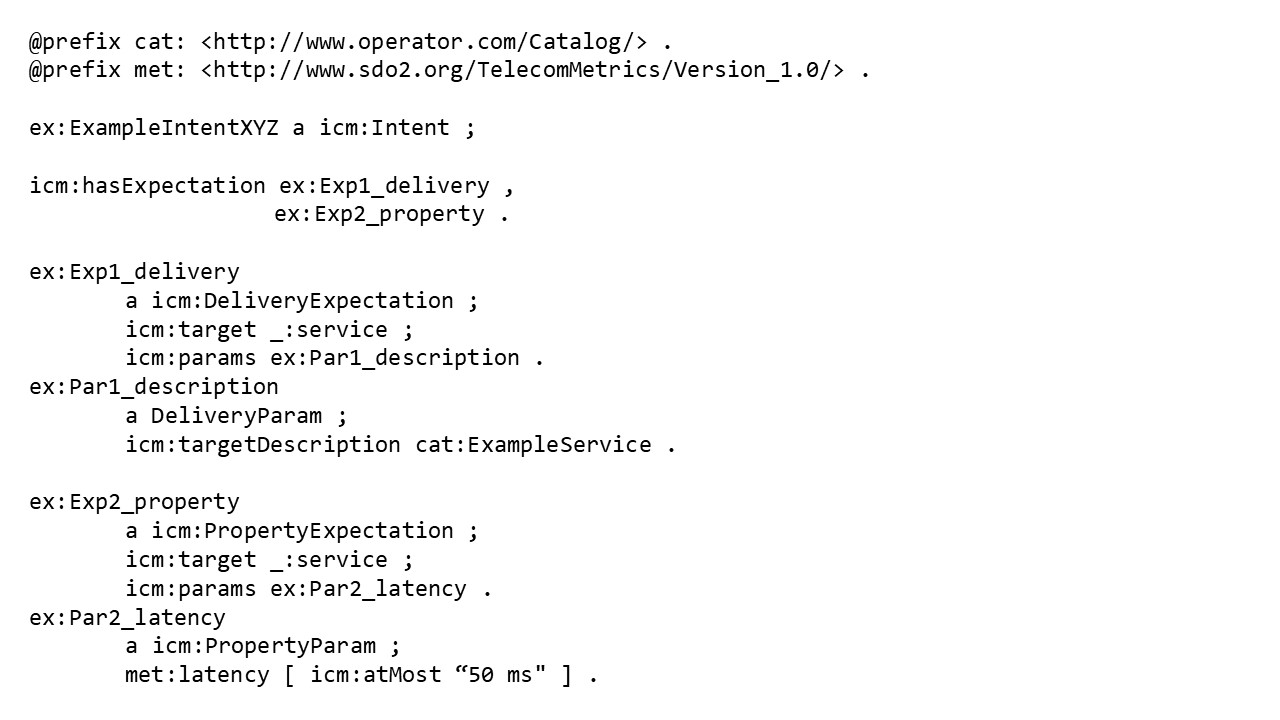}
        \end{center}
        \caption{Latency Metric Example}
        \label{fig:latency}
\end{figure}

When the drone crosses the boundary between edge sites, any of its microservices running on the source edge site should be migrated to the target edge site within specified latency limits, ensuring coordinated handover at application and network layers. This requirement would therefore be specified as an intent by the BSS layer to the SMO. A simple example of a latency metric, defined as per the intent common model proposed by TMForum, is shown in Fig.~\ref{fig:latency}. When it comes to successful handover for intent management function-based services, the UE context transfer/retrieval and bearer setup in the target should happen within a specified time. Thus, the Control Unit-Control Plane (CU-CP), Control Unit-User Plane (CU-UP), and Distributed Unit (DU) would have to coordinate to make this happen.

\begin{figure}
        \begin{center}
                \includegraphics[scale=0.42]{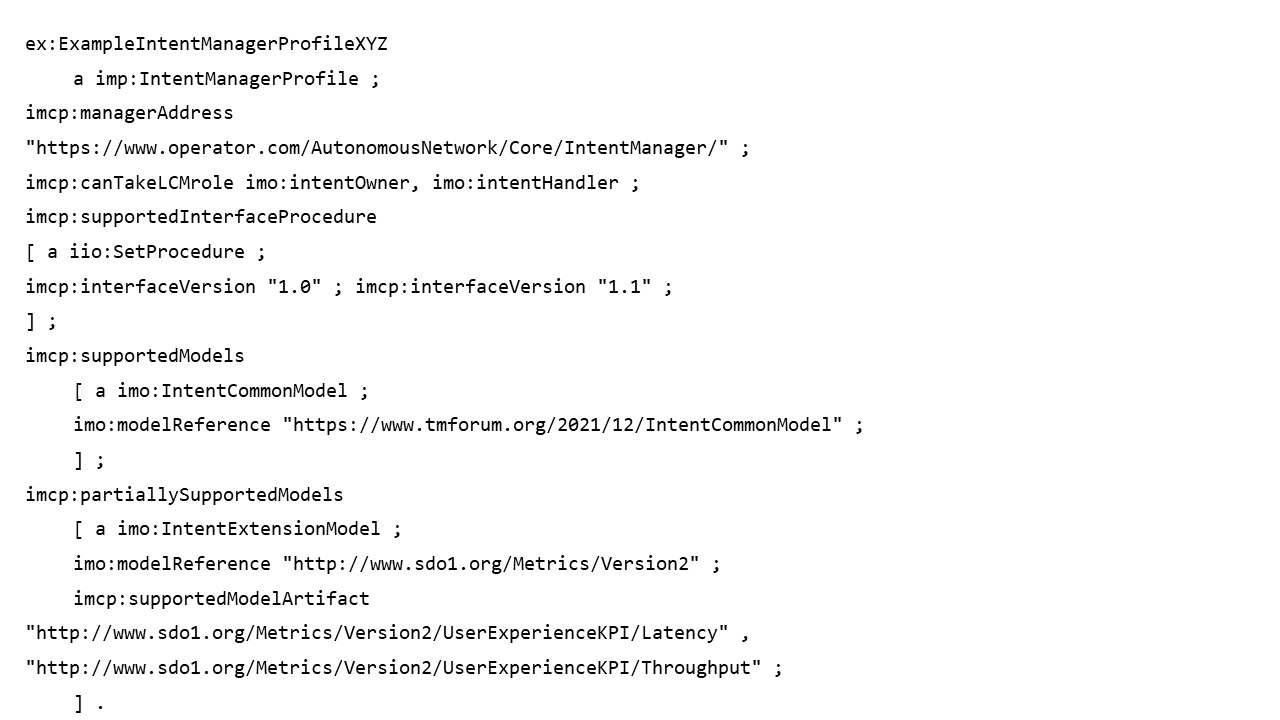}
        \end{center}
        \caption{Capability Profile Example}
        \label{fig:capability}
\end{figure}

Conversely, any IMF could advertise its capability via its capability profile~\cite{capability}, which would allow IMFs at the next higher layer to determine what type of intents it can handle. The capability profile of an IMF that can serve as both an intent owner and intent handler, and which can handle latency and throughput intents, is shown in Fig.~\ref{fig:capability}.

Based on the upper limit of 50 ms specified in Fig.~\ref{fig:latency}, and referring to Fig.~\ref{fig:figure-02}, the Service Management and Orchestration framework (SMO) - through the nonRT-RIC - would decompose this intent into, say, 20 ms for the cloud-native network functions in the O-Cloud located at the edge site (for the transport network, functions such as routers, firewalls, etc.), and 30 ms for the nearRT-RIC. The nearRT-RIC would decompose this further into suitable RRM intents at the CU-CP, with emphasis on Radio Resource Control (RRC). The CU will then decompose the intent into one or more DUs, with emphasis on Radio Link Control (RLC) and Medium Access Control (MAC). For example, one DU could take up 18 ms while the other DU could take up 12 ms. Finally, each DU will then implement its intent at its Radio Units (RU).

Please note that for concreteness and as an illustration, we have only described a rather simple latency metric example. More complex examples would involve setting a time duration within which a certain percentage of handovers should be implemented within a certain latency limit, e.g., 85\% of handovers within the next 30 minutes should be implemented within 50 ms. This would make radio resource allocation and scheduling at the RUs dynamic, requiring methods such as multi-agent online learning~\cite{hsieh2022multi} to fulfill the intent.

\section{Need for Meta-Scheduling}\label{sec:meta}


\subsection{Rapid Resource Allocation and Scheduling}\label{subsec:rapid}


Rapid (typically sub-millisecond) resource allocation and scheduling in programmable networks is crucial for several reasons:
\begin{itemize}
 \item Quality of Service (QoS): Programmable networks need to maintain specific QoS parameters. Rapid scheduling helps prioritize critical traffic and ensures that service level agreements (SLAs) are met.
 \item Network Slicing: With the emergence of 5G and eventually 6G, network slicing~\cite{thantharate2023adaptive6g} allows different applications to run on the same physical infrastructure while meeting diverse performance requirements. Rapid resource allocation is essential for managing these slices effectively.
 \item Latency Sensitivity: Many applications, such as real-time communications, online gaming, and UAVs (as evidenced from the example in Section~\ref{sec:resalloc} above), require extremely low latency. Delays in resource allocation can lead to degraded user experiences.
 \end{itemize}

In addition to the above, in large-scale real-life programmable networks, catering to multiple user requests would require multiple resource allocation and scheduling algorithms to be implemented, on the same network resources, leading to potential conflicts, which would degrade network performance considerably~\cite{corici2024towards,skaperas2021scheduling}. Preventing such conflicts from arising, would require a higher-level entity that coordinates all resource scheduling implementations, and in essence ensures co-existence of various scheduling algorithms for various use case types~\cite{kumar2023qos}.



\subsection{Meta Scheduling Architectural Framework}\label{subsec:meta}

\begin{figure*}
        \begin{center}
                \includegraphics[scale=0.55]{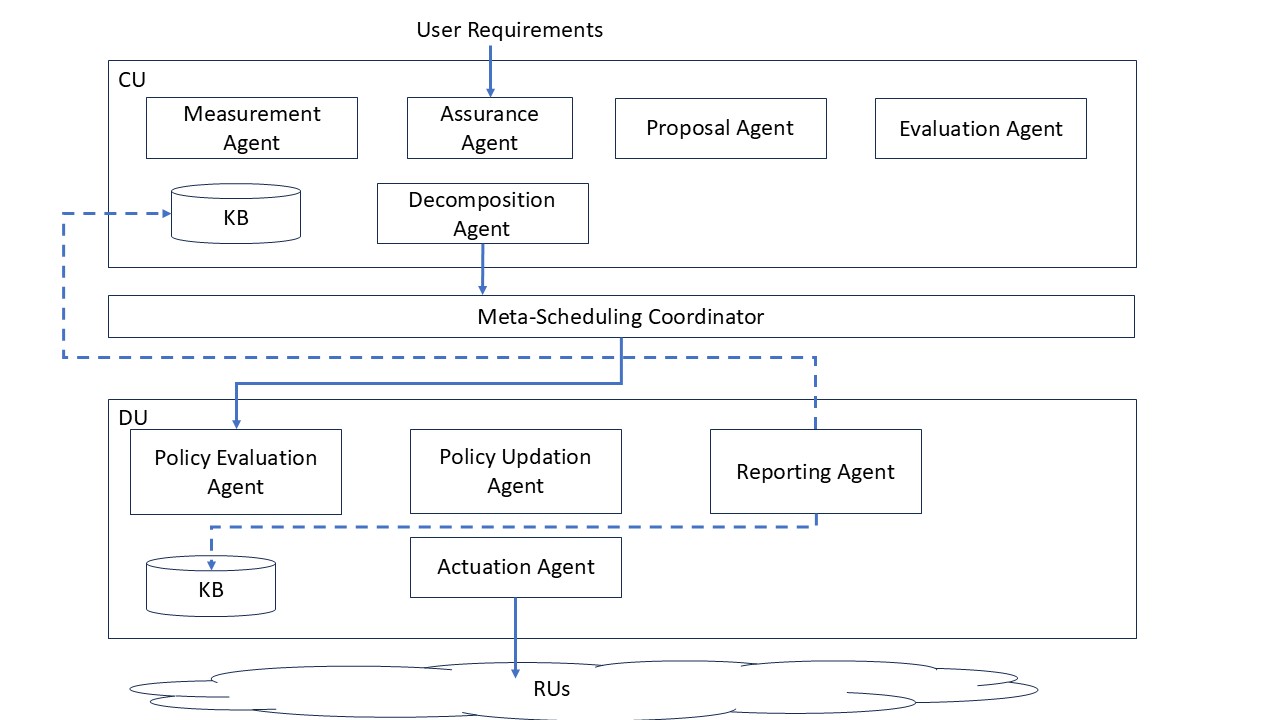}
        \end{center}
        \caption{Meta-Scheduling Architectural Framework}
        \label{fig:architecture}
\end{figure*}

Building on the ideas presented above, our meta scheduling architectural framework would therefore be a two-level framework that mirrors the hierarchy shown in Fig.~\ref{fig:jorg}. Applied to the O-RAN architecture of Fig.~\ref{fig:figure-02}, our framework would be as depicted in Fig.~\ref{fig:architecture}.

This framework would operate on two levels: meta-scheduling and scheduling. At the meta-scheduling level, the CU would be enhanced with the agents as depicted in Fig.~\ref{fig:architecture}, and would work as follows:
\begin{enumerate}
    \item The Assurance Agent would receive the user requirements. It is assumed that, due to the scale involved, multiple user requirements would come in at the same time. Referring to our UAV example above, this can be reflected in a need to schedule handovers for multiple UAVs at the same time, with differing latency requirements for each UAV.
    \item The Assurance Agent would evaluate the requirements with the help of the Measurement Agent. The latter would provide the former with information regarding the latest state of the network as recorded in the Knowledge Base (KB), as well as the underlying causal models that represent the variables that would affect scheduling decisions. These models are derived via \emph{active inference}, a type of causal inference that will be illustrated later in Section~\ref{subsec:active}. 
    \item The Proposal Agent would then develop meta-scheduling proposals (i.e., policies) to fulfill all the user requirements together while ensuring a fair (proportional or otherwise~\cite{prado2023enabling}) and conflict-free meta-scheduling approach. 
    \item The Evaluation Agent would evaluate the policies developed by the Proposal Agent and select the best policies that would meet the user requirements.
    \item The Decomposition Agent would send the selected meta-scheduling policy to the Meta-Scheduling Coordinator, for further forwarding to the various DUs. The appropriate scheduling policy to be assigned to each DU would be determined by the Decomposition Agent, and would be implemented as per the intent decomposition approach described earlier in Section~\ref{sec:resalloc}.
    \item The Meta-Scheduling Coordinator would serve as a message bus that transmits messages (intent decompositions) from the CU to DU, and responses (intent reports) from DU to CU.
\end{enumerate}

At the scheduling level, the DU would need to be enhanced thus:
\begin{enumerate}
    \item The Policy Evaluation Agent receives the scheduling policy via the Meta-Scheduling Coordinator. It then evaluates the given policy against its own KB to determine how feasible the policy would be, given the state of the network under its control as recorded in its own KB.
    \item The Policy Updation Agent would then update the scheduling policy as per the inputs from the Policy Evaluation Agent and send it to the Actuation Agent for action.
    \item The Actuation Agent would then implement the scheduling policy on the RUs. 
    \item The Reporting Agent would observe the results of scheduling and send its reports to the CU's KB (and as needed, to the DU's own KB); this would be implemented as per TMForum's Intent Reporting API~\cite{intent-reporting}. These intent reports would then be used to enhance the knowledge in both KBs, with a view towards improving meta-scheduling and scheduling algorithms in the future.
\end{enumerate}

The question now arises as to why the facility of policy updation at the DU level should be provided at all. The reason for this, is that this is in keeping with the TMForum's philosophy (as pictorially depicted in Fig.~\ref{fig:jorg}) of providing autonomy to intent management functions at every layer of the network stack. This is also in line with the recent proposal to complement rApps and xApps in the O-RAN standards~\cite{o-ran-standards} with ``dApps''~\cite{d2022dapps} at the DU, which can operate at $<$ 10 ms timescales and can be situated at the DU to perform scheduling.

By way of illustration, we have depicted in Algorithm~\ref{algo:algo1} how the meta-scheduler can help fulfill intents at the base station (gNB), using RAN schedulers for a given UE and the PDU session.



\begin{algorithm}
\caption{Meta Scheduling Algorithm}\label{algo:algo1}
\begin{algorithmic}[1]
\If {Intent}
\State CU and Meta Scheduler gets intent from IMF (Intent Management Function)
\State CU chooses the DRB based on the mapping function and DU uses the scheduling policy which is output of the meta scheduler to meet the intent within the same slice for differentiating the intent\\
DRB =  f(Intent, 5QI) \\
Meta Scheduling policy = g(Intent, Slice differentiator, Buffer status, CQI, Block Error rate)
\State RAN scheduling policy = h(Meta scheduling policy, Buffer status, CQI, Block error rate (BLER))
\Else \Comment {Intent is zero} \\
DRB =  f(0 , 5QI)  \\
Meta Scheduling policy = g(0, Slice differentiator, Buffer status, CQI, Block Error rate)
\State RAN scheduling policy = h(Meta scheduling policy, Buffer status, CQI, Block error rate (BLER))
\State when UE moves from one CU/DU to another CU/DU then the meta scheduler changes its inputs to the corresponding new scheduler, accordingly.
\State The input to the RAN scheduler is based on the data the meta scheduler has about the UE, the UE mobility and gNB (CU+DU).
\EndIf
\end{algorithmic}
\end{algorithm}

\subsection{Causal Reasoning for Scheduling}\label{subsec:causalreasoning}

Recent attempts at resource scheduling in 5G and B5G (beyond 5G) networks has focused on machine learning methods built on statistical principles. However, these methods suffer from several shortcomings as highlighted in~\cite{thomas2024causal}, viz., black box nature, curve fitting nature that limits their adaptability, reliance on large amounts of data, and energy inefficiency. Indeed, one key issue in adopting such machine learning approaches is model drift~\cite{10163748}, i.e., the fact that network conditions keep changing constantly, and hence the data on which the machine learning algorithms are run, would be quite dissimilar in distribution to the data on which the algorithms were originally trained.

This raises the need for a causal reasoning~\cite{pearl2016causal} approach, built on causal models that capture the relationships among the data in the network, and can use these model to enhance machine learning techniques used for resource scheduling. Causal models for any variable in the network can be further refined via the use of \emph{Markov Blankets}~\cite{kirchhoff2018markov}. In a causal model, which is represented as a directed acyclic graph (DAG), the Markov Blanket of any variable is the collection of its parents, children and co-parents in the DAG. The Markov Blanket would therefore comprise those variables that would affect the variable in question. Hence any learning algorithm that seeks to determine the value of the variable would need only consider the members of the Markov Blanket as independent variables. The Markov Blanket for any variable can be discovered via techniques such as those described in~\cite{tsamardinos2003algorithms}.

\begin{figure}
        \begin{center}
                \includegraphics[scale=0.27]{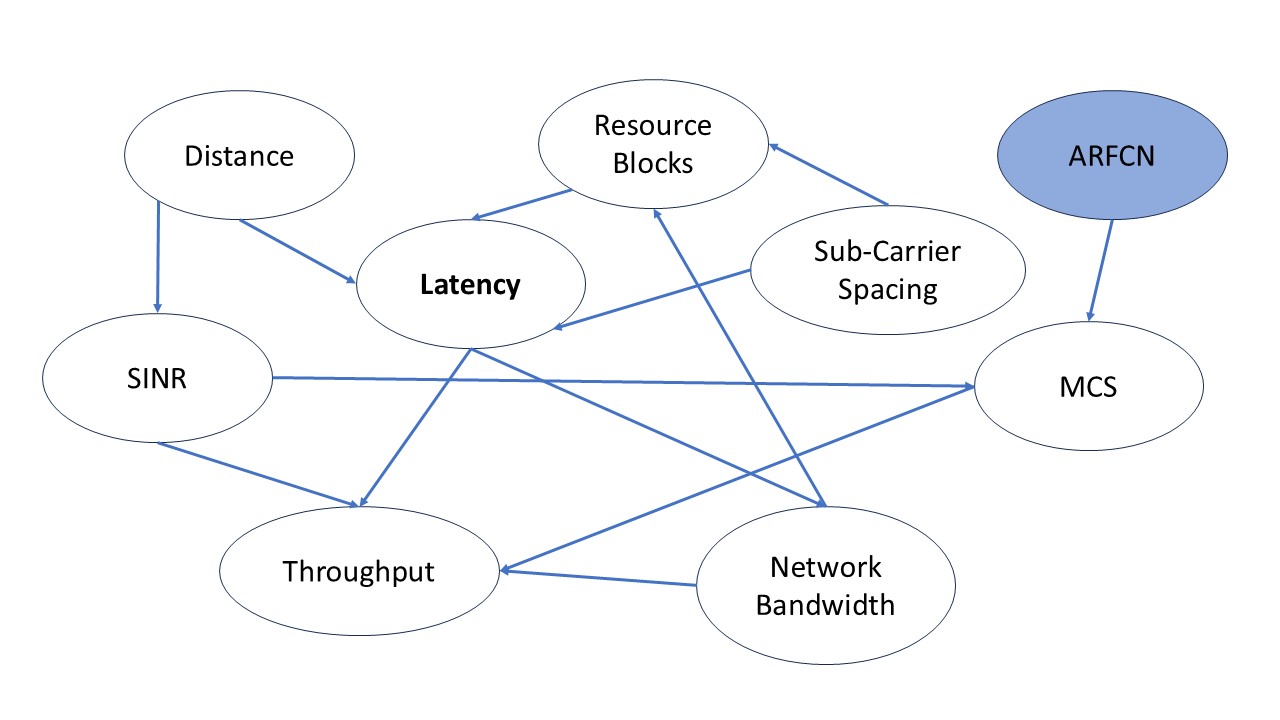}
        \end{center}
        \caption{Factors that could affect Latency}
        \label{fig:markov}
\end{figure}

As a simple illustration, consider Fig.~\ref{fig:markov}, which shows some (not all) factors that could affect latency. The factor ARFCN is the exception in Fig.~\ref{fig:markov}, and is shaded since it does not belong to the Markov Blanket of latency.

Techniques such as those described in~\cite{sousa2024enhancing} could be developed to uncover causal relationships among the variables in the CU, DU and RU of the O-RAN. Indeed, the technique in~\cite{sousa2024enhancing}, used unsupervised learning to uncover a causal association between a Configuration Management parameter and degraded performance of a set of Base Stations (BSs). The technique used was an autoencoder based on an unsupervised Deep Neural Network (DNN), which extracted a lower-dimensional representation from the Performance Management (PM) and CM indicators of each BS, simplifying the subsequent application of clustering
algorithms. The clustering algorithms were used to group the BSs with similar performance as per their PM values.

\subsection{Extending Causal Reasoning with Active Inference}\label{subsec:active}

Active inference~\cite{sedlak2024active} is an extension of causal inference. It is a concept originally from neuroscience which models how the brain constantly predicts and evaluates sensory information to decrease long-term surprise. ``Surprise'' of any observation given a model is modeled as the negative log-likelihood of the observation. Surprise is typically defined as so-called ``Free Energy'' (FE), which is the gap between any observer's understanding and the reality. The FE is usually modeled via the Kullkack-Leibler (KL) divergence $D_{KL}$ between approximate posterior probability ($Q$) of hidden states ($x$) and their exact posterior probability ($P$) (as shown in Equations~\ref{eq1} and~\ref{eq2}) reproduced from~\cite{sedlak2024active}. 

\begin{equation}\label{eq1}
    \Im(o|m) = -ln~P(o|m)
\end{equation}

\begin{equation}\label{eq2}
    F[Q ,o] = D_{KL}[Q(x)||P(x|o,m)] + \Im(o|m) \geq \Im(o|m)
\end{equation}

\begin{figure}
        \begin{center}
                \includegraphics[scale=0.28]{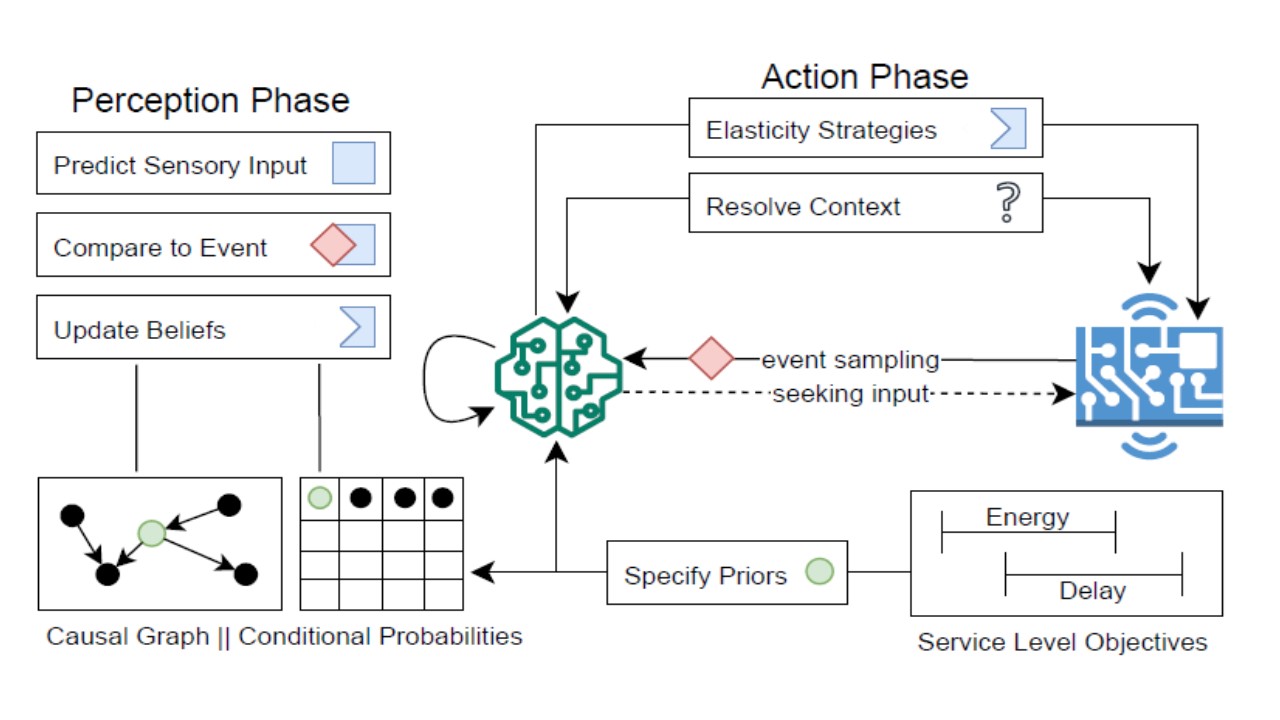}
        \end{center}
        \caption{Action Perception Cycle for Active Inference Agent - from~\cite{sedlak2024active}}
        \label{fig:activeinference}
\end{figure}

Active inference agents work on action-perception cycles, where (a) they predict the outcomes of their actions based on their beliefs, and (b) update their beliefs based on the results of their actions. This works as depicted in Fig.~\ref{fig:activeinference}. First, the agent is given a set of expectations that it needs to meet, for e.g., in our case, latency. The agent creates a causal model (e.g., Fig.~\ref{fig:markov}) to determine the factors that influence the expectation. This is represented as a conditional probability table that contains the degree to which the factors influence the expectation in question. After this, the agent starts to continuously evaluate the event against the expectation. To decrease the Free Energy, the agent can: (1) adjust its beliefs accordingly; (2) execute elasticity strategies; or (3) resolve the contextual information to improve decision-making.

In our proposed meta-scheduling architecture as depicted in Fig.~\ref{fig:architecture}, both prediction and belief updation would be accomplished in the KBs of the CU and DU. Actions would be implemented via the Decomposition and Actuation Agents, while results of the actions would be obtained via the Reporting Agent.

Referring to Fig.~\ref{fig:markov}, active inference-related actions could therefore be limited to modifying variables such as distance, Resource Block allocations, and Modulation and Coding Schemes, in order to achieve the desired latency. And in the perception phase of the action-perception cycle, the beliefs (typically expressed as Bayesian probabilities) associated with these variables would need to be adjusted based on the KL divergence between the planned and actual latency values.

\begin{figure}
        \begin{center}
                \includegraphics[scale=0.32]{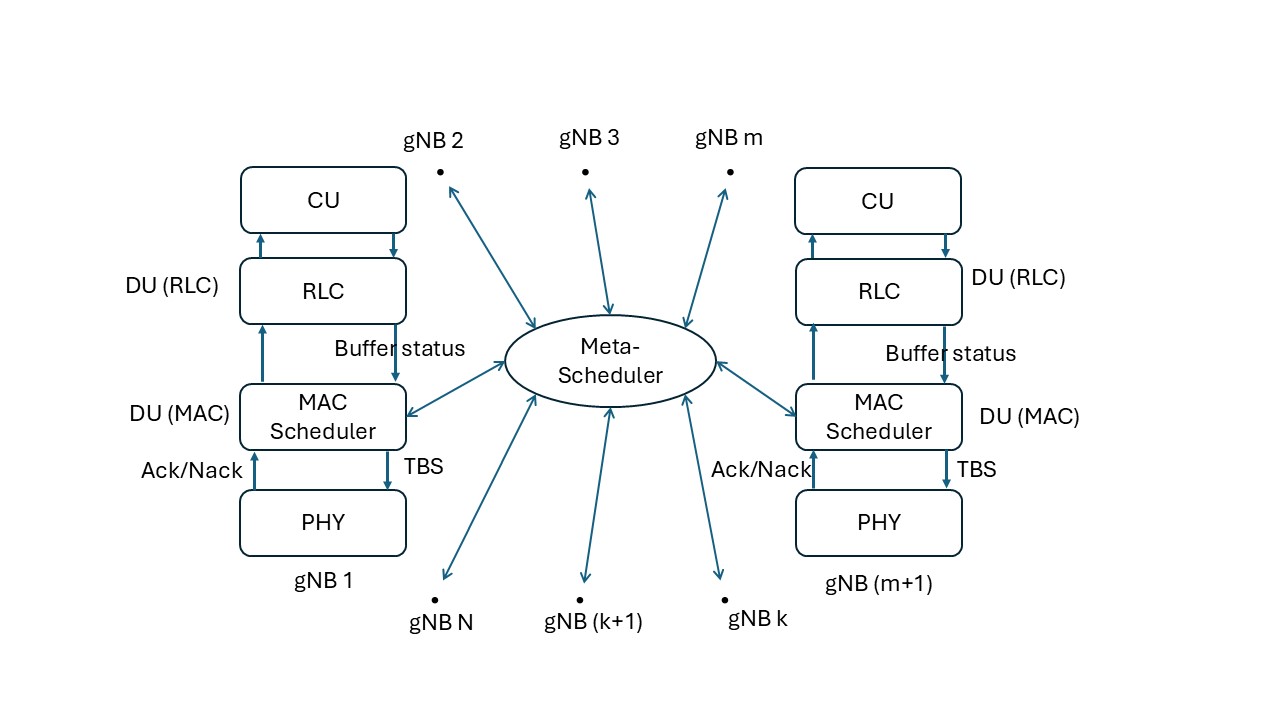}
        \end{center}
        \caption{Working of Meta-scheduler with Schedulers}
        \label{fig:meta-scheduler-working}
\end{figure}

Scheduling decisions in 5G base stations (gNBs) usually depend on factors such as Radio Link Control (RLC) buffer status, Block Error Rate (BLER) obtained by ACK or NACK received from the physical (PHY) layer, and the actual scheduling mechanism such as round robin or proportional fair. These factors will help select the user to be served. After that, the scheduler collects the buffered data from RLC and sends it to PHY for transmission. This is pictorially depicted in Fig.~\ref{fig:meta-scheduler-working}.


Once active inference is integrated into our architectural framework, it can be used to speed up machine learning algorithms used for resource scheduling and also make them more accurate by only providing them data of the independent variables in the Markov Blankets of the dependent variables which need to be optimized. Some recent examples of prior work that can be considered, are:
\begin{itemize}
    \item Meta-scheduling with cooperative learning~\cite{min2023meta}: a two-layer meta-scheduling (at the CU level) and scheduling framework (at the DU level) that uses cooperative learning between the two levels to optimize scheduling. It proposes the use of Deep Reinforcement Learning at the scheduling level, while meta-scheduling is implemented via a meta-RL policy. In particular, the meta-RL policy addressing problems with an identical task having different system dynamics, as presented in~\cite{lee2022system} is employed. 
    \item An example of an intent-driven closed loop management for 6G O-RAN is presented in~\cite{zhang2023intent}. Its focus, however, is on automating network slice management, although it provides an intent decomposition model that can be considered for incorporation into our meta-scheduling framework. However, it only presents a single-layer scheduling approach involving Deep Q-Learning, and does not consider causal inference.
    \item It is increasingly seen in most wireless networks that traditional offline learning approaches cannot adapt to rapid changes in network conditions, which would be expected in 5G/B5G networks. To that end, online learning techniques, in particular, multi-armed bandits are becoming popular. One such example is presented in~\cite{karakaya2024online}, which describes a hierarchical multi-armed bandit technique for effective intent-based management. Similar to our proposed two-layer framework,~\cite{karakaya2024online} proposes a two-layer closed loop framework, where child agents in the bottom layer are assigned specific intent key performance indicators (KPIs) that they should meet. The child agent then selects an action which is then evaluated by the parent agent and the action with least pseudo-regret (which quantifies the difference between the expected reward
    achieved by the optimal and selected arms) is selected by the parent agent. This, however, differs from our approach in two ways. First, in our approach the CU at the meta-scheduling layer would send a scheduling policy along with the intent KPI, and the DU would be free to evaluate and accept or modify it as per its situation. Second, there is no concept of an action-perception cycle in~\cite{karakaya2024online}.
    \item An intent-driven orchestration method of cognitive autonomous networks (CANs) for RAN management is presented in~\cite{banerjee2021intent}. That paper contains a high-level description of an end-to-end architecture for for intent-driven management of RAN parameters within the CAN. It introduces the concepts of Intent Specification Platform, an Intent Fulfillment System, and an Intent-driven Network Automation Function Orchestrator (IDNAFO). When  our meta-scheduling framework is to be implemented and demonstrated, these three concepts can be incorporated into it. 
\end{itemize}

\section{Key Research Questions}\label{sec:key}


Based on the above discussion, we identify the following key research questions. This list is not necessarily exhaustive, and we believe it would expand as the research questions themselves begin to be investigated:
\begin{itemize}
    \item Modeling-related: related to modeling the meta-scheduling architectural framework, and the inference algorithms for meta-scheduling and scheduling:
    \begin{enumerate}
        \item Methods for intent decomposition and assigning the appropriate scheduling policies via the Meta-Scheduling Coordinator. Intent decomposition could be built on techniques such as those proposed in works such as~\cite{christou, zhang2023intent,gritli,wang2024networkintentdecompositionoptimization,DBLP:conf/aiml2/KattepurConrad}.
        \item Causal model discovery in programmable networks, perhaps building on works such as~\cite{sousa2023towards}. One special issue to contend with here, would be the size and scale of the programmable network itself, which is expected to be highly distributed and composed of several administrative domains, which may impact each other, especially at the time of intent decomposition, as shown in~\cite{christou}. As far as we are aware, this problem remains unsolved.
        \item Elasticity strategies as proposed in~\cite{sedlak2024active} to maintain homeostasis, i.e., persistence of adherence to user requirements over time. This is crucial to ensure the continual adherence of the network to user requirements, especially since such requirements are expected to be dynamic. 
        \item Integration of active inference into any machine learning algorithms employed to perform resource scheduling. We expect that the meta-scheduling layer in the CU would need to handle multiple such algorithms being implemented at the same time, and they would be heterogeneous. This heterogeneity would therefore require special techniques to optimize meta-scheduling using active inference over large numbers of instances of scheduling implementations. This would also tie into the above point of maintaining homeostasis in the midst of such heterogeneity. Techniques from causal machine learning~\cite{kaddour2206causal,roy2024causalitydrivenreinforcementlearningjoint} would need to be investigated here.
        \item Since O-RAN is expected to be a key model for programmable networks going forward to 5G and beyond, several enhancements to O-RAN standards~\cite{o-ran-standards} would need to be investigated, which include the following: intent-based management; O1 interface for Orchestration and Management (O-RAN interfaces are depicted in Fig.~\ref{fig:figure-02}); E2 interface between nearRT-RIC and O-DU; as well as Fronthaul interfaces between O-DU and O-RU. All these interfaces would need to be enhanced to incorporate our meta-scheduling framework, including the meta-scheduling layer at the CU, Meta-Scheduling Coordinator that would exercise the E2 interface, and the scheduling layer where the Fronthaul interfaces would need to be incorporated.
    \end{enumerate}
    \item Implementation-related: related to implementation issues facing the meta-scheduling framework:
    \begin{enumerate}
        \item How to actually implement the framework on large-scale realistic 5G/B5G use cases, such as UAVs, Vehicle-to-Everything (V2X)~\cite{alalewi20215g}, high-traffic urban networks in smart city deployments~\cite{ali2023enabling}, and possibly non-terrestrial deployments~\cite{majamaa2024toward} as well.
        \item How to address operational challenges and failure scenarios, and how dynamic meta-scheduling can help build resilience into resource scheduling~\cite{de2024performance}.
    \end{enumerate}
\end{itemize}


\section{Conclusions}

In this position paper, we have introduced the key research issue of managing programmable networks. In particular, we have highlighted the problem of intent-based meta-scheduling in such networks. We have shown how the principles of active inference, derived from the well-known idea of causal inference, can be used to model and manage a two-layer meta-scheduling framework that can separate out the overall task of managing the network in line with overall user requirements, from the individual tasks of resource scheduling for meeting specific intent KPIs. We concluded our paper by presenting key research questions that need to be addressed in order to make intent-based meta-scheduling a reality.

\bibliographystyle{IEEEtran}
\bibliography{refs}

\begin{appendices}
\section{Intent Decomposition Method from~\cite{christou}}\label{sec:app-a}

\subsection{Description of~\cite{christou}}\label{subsec:description}

\begin{figure}[htbp]
        \begin{center}
                \includegraphics[scale=0.28]{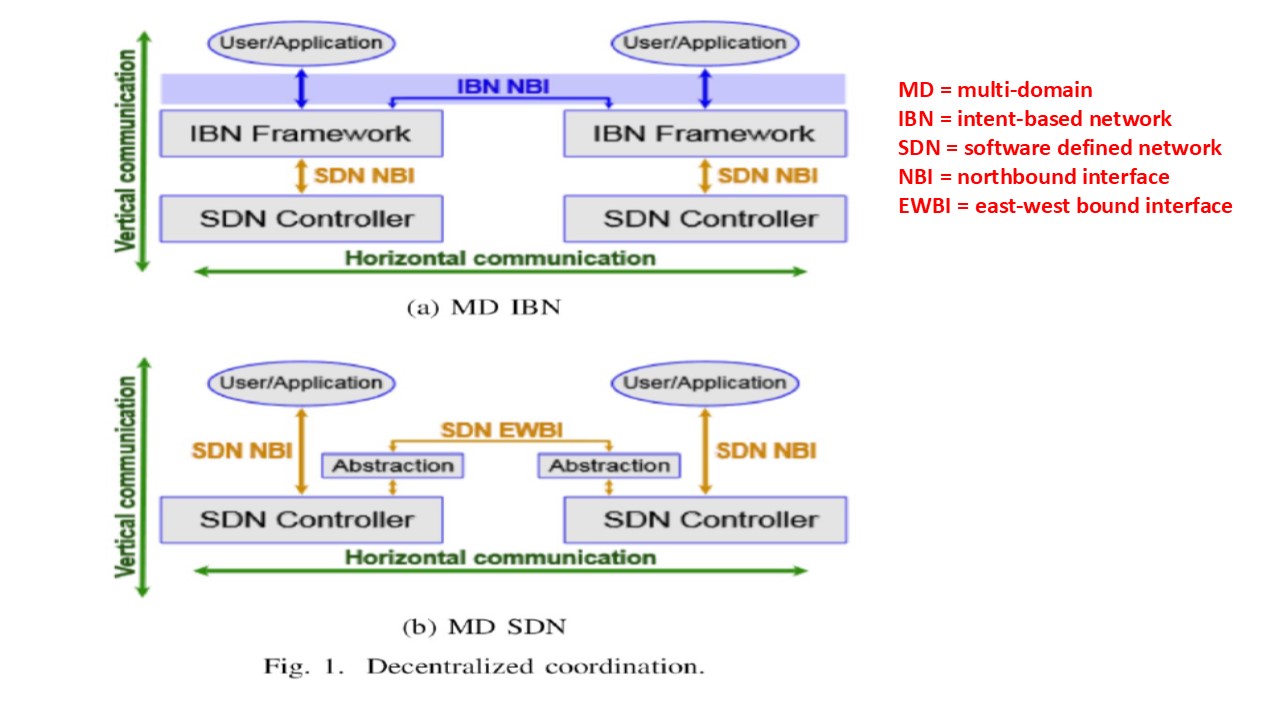}
        \end{center}
        \caption{Decentralized Coordination of Intent-based Networks - from~\cite{christou}}
        \label{fig:ibn}
\end{figure}

The intent decomposition approach in~\cite{christou} has been developed for IP-optical networks, although its approach is general enough for any 5G/B5G wireless network. The emphasis of the approach in~\cite{christou} is decentralized coordination using multiple SDN controllers, as depicted in Fig.~\ref{fig:ibn}.

\begin{figure}[htbp]
        \begin{center}
                \includegraphics[scale=0.28]{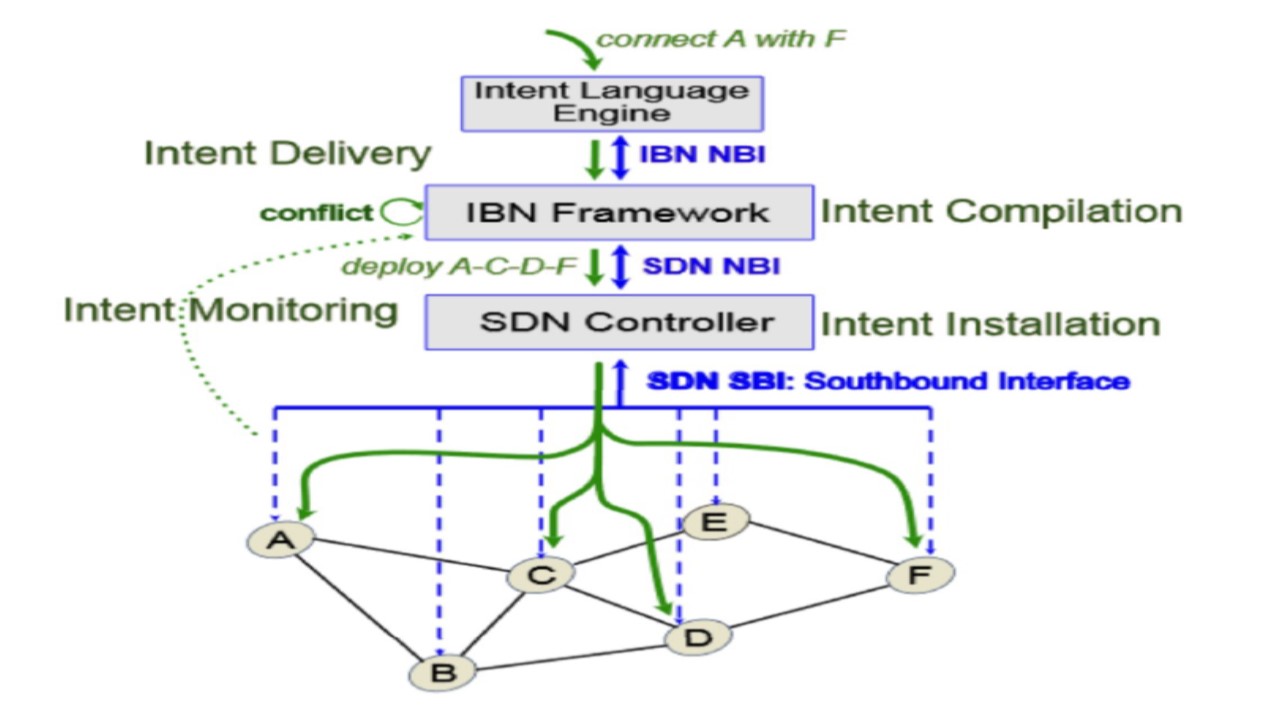}
        \end{center}
        \caption{IBN over SDN architecture - from~\cite{christou}}
        \label{fig:sdn}
\end{figure}

Fig.~\ref{fig:sdn} illustrates the various intent stages as per~\cite{christou}. First the intent enters the system expressed in an intent language. The intent language engine uses the IBN NBI to insert the intent into the IBN framework (Intent Delivery). The IBN framework processes the intent, generates a potential implementation (Intent Compilation), and forwards it to the SDN Controller to be deployed in the required devices (Intent Installation). The performance of intent fulfillment is continuously monitored (Intent Monitoring). Any conflict arising out of satisfying multiple intents at the same time should also be addressed, although that is outside the scope of~\cite{christou}.

\begin{figure}[htbp]
        \begin{center}
                \includegraphics[scale=0.28]{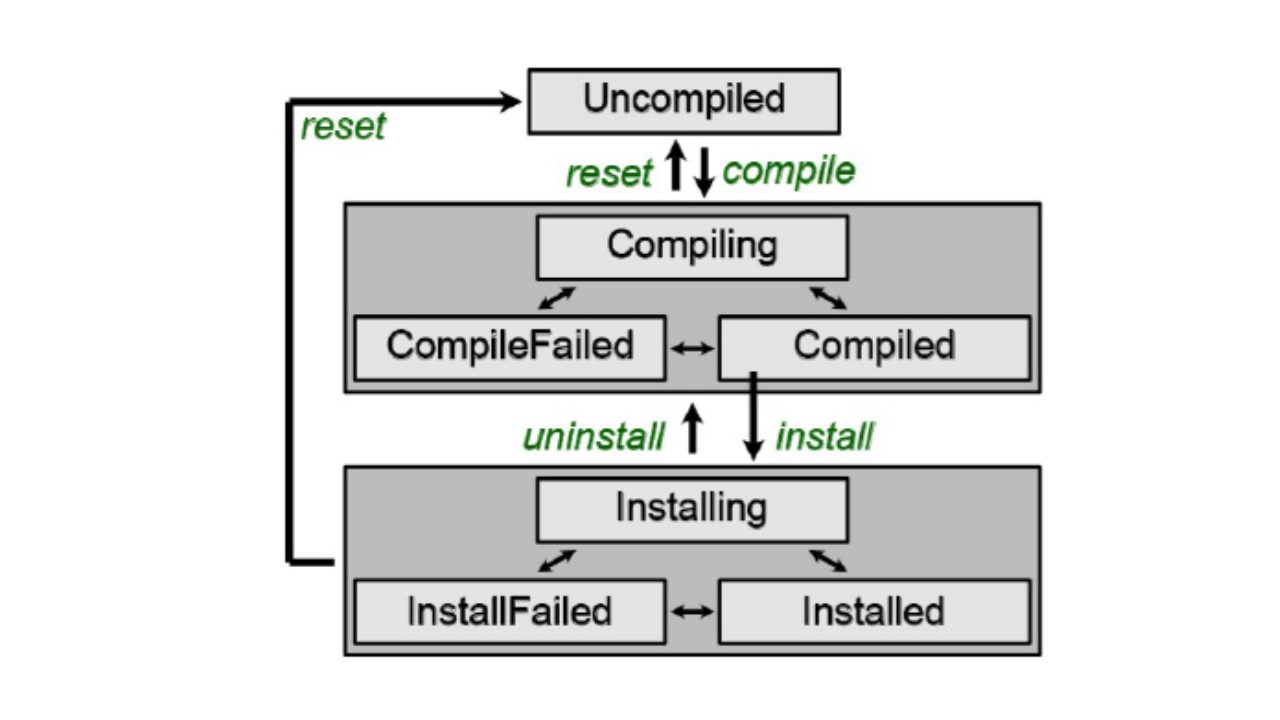}
        \end{center}
        \caption{Intent State Machine - from~\cite{christou}}
        \label{fig:intentstatemachine}
\end{figure}

The intent state machine is depicted in Fig.~\ref{fig:intentstatemachine}. To get installed, an intent must first be compiled. \emph{Compiling} and \emph{Installing} are intermediate steps that signify dependence on the child intents in the intent tree. Compiling and installation will fail if resources are unavailable or if the intent requirements are not satisfied. 

\begin{figure*}[t]
        \begin{center}
                \includegraphics[scale=0.55]{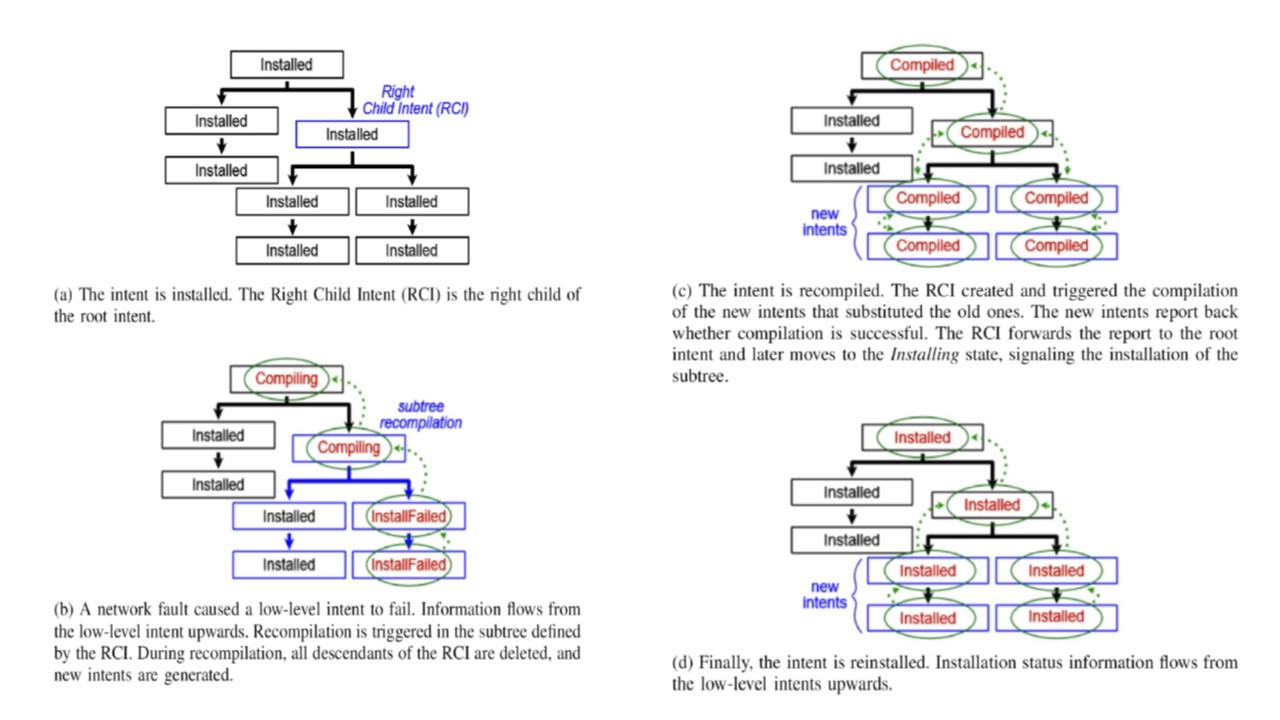}
        \end{center}
        \caption{Intent State Propagation in case of a Network Fault - from~\cite{christou}}
        \label{fig:intentstatepropagation}
\end{figure*}

Hence the work in~\cite{christou} is based on the concept of \emph{intent tree}, whose root is the received intent. Each intent can be broken down into sub-intents and can be considered installed when all children are all installed. This triggers updates to the parent's state based on the child's state. Failure states of children are propagated to the ancestors, who can decide to take the appropriate actions, i.e., try to address the failure or recompile the intent. This process is depicted in more detail in Fig.~\ref{fig:intentstatepropagation}. 

The paper~\cite{christou} considers only best-effort connectivity intents, with correspond to the Routing and Spectrum Assignment (RSA) problem~\cite{chatterjee2015routing}. Since RSA is NP-Hard, it is split into (1) routing and (2) spectrum allocation subproblems. The strategy employed in~\cite{christou} assigns a \emph{PathIntent} for every \emph{ConnectivityIntent} to solve the routing subproblem, and a \emph{SpectrumIntent} to solve the spectrum allocation subproblem.

Scaling to multi-domain networks (MD) is done via the use of a \emph{RemoteIntent}, which delegates an intent to another domain by binding the local intent to a new replica on the remote domain with a parent-child relationship. The state update properties still hold here like any parent-child relationship in the intent tree. This way, the intent states can propagate across multiple administrative domains. 

\begin{figure*}[htbp]
        \begin{center}
                \includegraphics[scale=0.55]{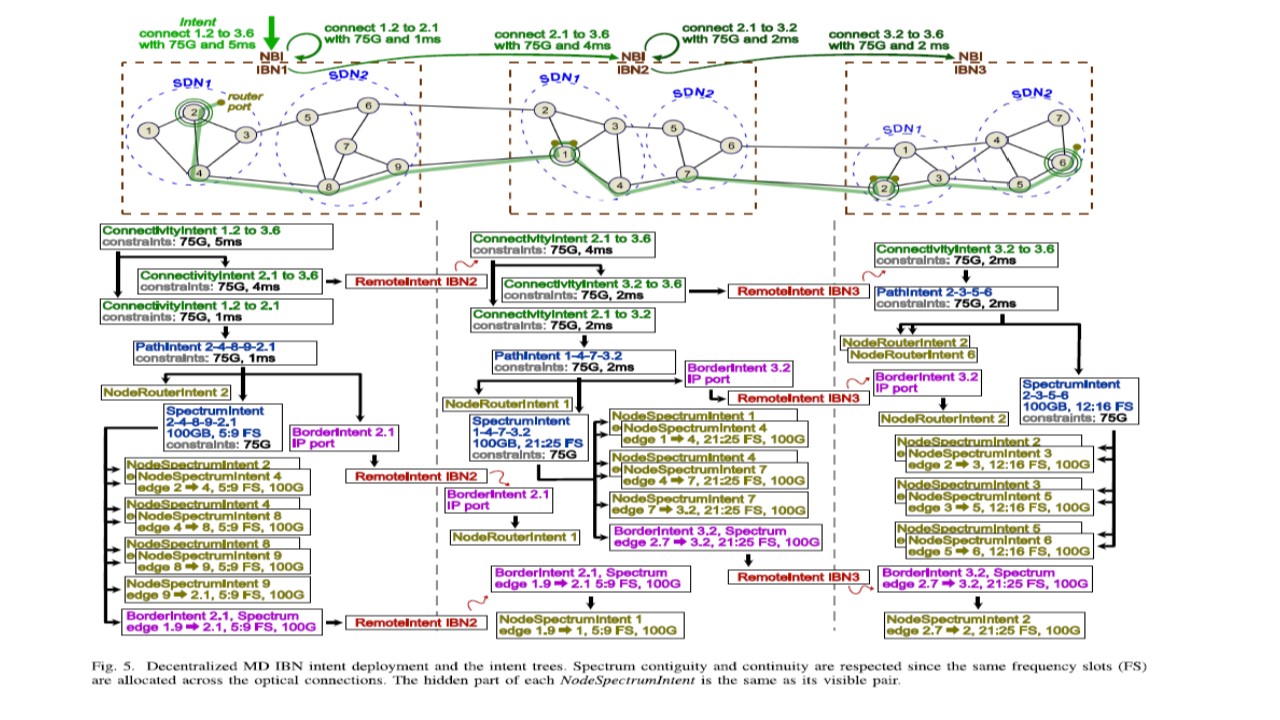}
        \end{center}
        \caption{Multi-domain intent deployment and the intent tree - from~\cite{christou}}
        \label{fig:md-deployment}
\end{figure*}

Fig.~\ref{fig:md-deployment} illustrates a prototype implementation of the above ideas, where the intent trees are generated while issuing to IBN1 a MD \emph{ConnectivityIntent} between nodes 1:2 and 3:6 with 5 ms latency and 75 Gbps bandwidth requirements. Node x.y signifies the $y-th$ node of the $x-th$ IBN domain. Overall, the IBN1 compiled the intent by subdividing
it into two ConnectivityIntents, one implemented locally while the other delegated to the neighboring domain. The current intent compilation strategy performs signal regeneration in the IP layer at every border node, i. e., nodes 2:1 and 3:2. The selection of the border nodes and the neighboring domain is based on the specifics of the deployed implementation
algorithm, i. e., the operator’s decision-making process. 

It is observed that PathIntents and SpectrumIntents compile down to low-level intents, i.e., NodeRouterIntents requesting IP router ports and NodeSpectrumIntent pairs requesting fiber spectrum slots for each node participating in the link. However, the IBN instance cannot control the neighboring domain for the inter-domain links, and a BorderIntent is generated instead, creating remote low-level intents for the border nodes. For example, to use the link between 1:9 and 2:1, the frequency slots 5; 6; 7; 8; 9 must be allocated at nodes 1:9 and 2:1. IBN1 creates a NodeSpectrumIntent for the local node 1:9 and a BorderIntent that will issue a RemoteIntent to IBN2 for 2:1. 

It is also noticed that constraints are propagated altered to the child intents, depending on whether they are guaranteed to be already (partly) satisfied by the parents or not. For example, the latency constraint of 5 ms is propagated to one of the child intents as a constraint of 1 ms. This means the parent guarantees that the intent constraint of 5 ms will be satisfied as long as the child satisfies the intent constraint of 1 ms. The PathIntent can decide if the delay constraint is satisfied since it knows the path. If it is satisfied, there is no reason to propagate the constraint further down to the child intents. If it is not satisfied, then the intent state will transition to \emph{CompileFailed}. If it is generally unknown whether the constraint is satisfied, the intent will transfer the constraint to the child intents unaltered.

When all the IBN instances successfully compile and install the system-generated intents, the end-to-end (E2E) connection will be available. If one of the IBN instances does not stand up to the requirements of an intent, this will be spotted from the monitoring procedure, which will update the state of the corresponding intent to \emph{InstallFailed}, making it clear whom to hold responsible. Such monitoring promotes accountability and conformity with the intent requirements.

\subsection{Analysis of~\cite{christou}}\label{subsec:analysis}

The paper~\cite{christou} presents an overview of how intents could be decomposed, and how the decomposition can be managed to ensure intent fulfillment. While we have cited many other intent decomposition methods from the literature~\cite{wang2024networkintentdecompositionoptimization,gritli,zhang2023intent}, the key aspect of~\cite{christou} is its treatment of intent decomposition across multiple administrative domains, which would be a key feature of programmable networks.

\end{appendices}
\end{document}